\documentclass[a4paper]{article}
\usepackage{amsmath,graphicx}
\usepackage{booktabs,adjustbox,makecell,multirow}
\usepackage{enumitem}
\usepackage[labelformat=simple]{subcaption}

\usepackage[font=small,labelfont=bf]{caption}
\usepackage{flushend}
\usepackage{INTERSPEECH2021}

\usepackage[dvipsnames]{xcolor}

\usepackage{hyperref}
\usepackage{marvosym}

\title{Target-speaker Voice Activity Detection with Improved I-Vector Estimation for Unknown Number of Speaker}
\name{\begin{tabular}{c}Maokui He$^1$, Desh Raj$^2$, Zili Huang$^2$, Jun Du$^{1,*}$, Zhuo Chen$^3$, Shinji Watanabe$^2$\end{tabular}}
\address{$^1$University of Science and Technology of China, HeFei, China\\$^2$Center for Language and Speech Processing, The Johns Hopkins University, Baltimore, MD\\$^3$Microsoft Corp, Redmond, WA}
\email{{\href{mailto:jundu@ustc.edu.cn}{\Letter}}jundu@ustc.edu.cn}

\begin{document}

\maketitle
\begin{abstract}
Target-speaker voice activity detection (TS-VAD) has recently shown promising results for speaker diarization on highly overlapped speech. However, the original model requires a fixed (and known) number of speakers, which limits its application to real conversations. 
In this paper, we extend TS-VAD to speaker diarization with unknown numbers of speakers. This is achieved by two steps: first, an initial diarization system is applied for speaker number estimation, followed by TS-VAD network output masking according to this estimate.
We further investigate different diarization methods, including clustering-based and region proposal networks, for estimating the initial i-vectors. Since these systems have complementary strengths, we propose a fusion-based method to combine frame-level decisions from the systems for an improved initialization. We demonstrate through experiments on variants of the LibriCSS meeting corpus that our proposed approach can improve the DER by up to 50\% relative across varying numbers of speakers. This improvement also results in better downstream ASR performance approaching that using oracle segments. 
\end{abstract}

\section{Introduction}
\label{sec:intro}

Speaker diarization refers to the task of segmenting a given recording into homogeneous speaker-specific regions~\cite{Mir2012SpeakerDA, Tranter2006AnOO}. The conventional approach for diarization~\cite{GarciaRomero2017SpeakerDU, Sun2018SpeakerDW} involves applying speech activity detection (SAD) followed by clustering of fixed-dimensional speaker embeddings -- usually i-vectors~\cite{Dehak2011FrontEndFA} or neural embeddings~\cite{Variani2014DeepNN,Snyder2018XVectorsRD} --- extracted from small subsegments of the speech regions. This may optionally be followed by a resegmentation step~\cite{Bozonnet2010TheLR,Sell2015DiarizationRI}. However, this framework inherently makes a single speaker assumption, since every subsegment can only be assigned a single label through hard clustering.

There have been efforts to solve the overlap problem in speaker diarization, by leveraging a separate overlap detection module that identifies segments containing overlapped speech. This overlap detection may be done using hidden Markov models (HMMs)~\cite{boakye2008overlapped, Huijbregts2009SpeechOD} or using neural networks~\cite{Hagerer2017EnhancingLR, Kunesov2019DetectionOO}. Additional speaker labels may be assigned to the overlapping segments once detected~\cite{Bullock2019OverlapawareDR}. An alternate framework that is popular in recent years applies end-to-end systems trained with supervision (often using simulated mixtures). This includes models such as end to end neural diarization(EEND)~\cite{Fujita2020EndtoEndND} or region proposal networks (RPN)~\cite{Huang2020SpeakerDW}, which can naturally handle overlaps, and have shown promising results on challenging data sets.

The target-speaker voice activity detection (TS-VAD) system was proposed recently~\cite{Medennikov2020TargetSpeakerVA}, and demonstrated promising results in the CHiME-6 challenge~\cite{Watanabe2020CHiME6CT}. The model was inspired from advances in target speech extraction methods (such as SpeakerBeam~\cite{molkov2017SpeakerAwareNN}, VoiceFilter~\cite{Wang2019VoiceFilterTV}, and Personal-VAD~\cite{Ding2019PersonalVS}), which utilize the target speakers' enrollment information to estimate time-frequency masks that indicate their voice activity in the recording. Inspired from supervised diarization methods like EEND, the network generates multi-speaker outputs containing frame-wise posteriors by leveraging the conditional dependence of speakers in a recording. TS-VAD achieved the best diarization performance on CHiME-6 (which contains 34\% overlap duration in the evaluation set), improving the baseline agglomerative hierarchical clustering (AHC) system by over 30\% absolute diarization error rate (DER).

Nevertheless, TS-VAD is limited in that the model assumes a fixed number of speakers, which must be known \textit{a priori}. This is because the neural network is trained with a fixed number of output nodes, and each of these nodes corresponds to a different speaker's activity during inference. This assumption hinders its application on recordings with varying or unknown numbers of speakers, which commonly happen in natural meetings. Furthermore, the performance of TS-VAD largely depends on the initial estimate of the speaker i-vectors --- a poor initialization may lead to substantially worse performance and more iterations for convergence. 


To address these limitations, in this work, we propose to extend TS-VAD for processing long-form recordings with unknown number of speakers.
Our approach relies on using a separate diarization model (typically the same system used for the initial i-vector estimates) to predict the number of speakers in the recording, and consequently manipulates the fixed TS-VAD network outputs to correspond to these speakers. 
We also investigate the effectiveness of different initialization diarization systems --- such as clustering-based or RPNs --- for obtaining the initial estimate of speaker i-vectors. Since these systems usually contain complementary strengths, we then propose a weighted mean strategy to combine their frame-level decision to get an optimal initialization for the TS-VAD model. 
Through experiments conducted on LibriCSS dataset~\cite{Chen2020ContinuousSS} and its complementary 2 speaker and 5 speaker sets, we show that our proposed TS-VAD extension improves the DER by up to 50\% relative, compared with a strong clustering-based baseline system. This improvement is consistent across recordings containing different numbers of speakers, even when using the same model for inference, which demonstrates the robustness of our approach. Furthermore, our fusion-based initialization technique provides 3.9\% relative improvement over the single best initialization. 

We also report the effect of our TS-VAD when integrated in a meeting transcription system \cite{Raj2021MultiSC}, in terms of concatenated minimum-permutation word error rate (cpWER)~\cite{Watanabe2020CHiME6CT} on whole recordings. We find that using segments obtained from our best TS-VAD output is comparable with using oracle segments for ASR.
\section{TS-VAD for unknown number of speakers}
\label{sec:ts-vad}

\begin{figure}[t]
 \centering
 \includegraphics[width=\linewidth]{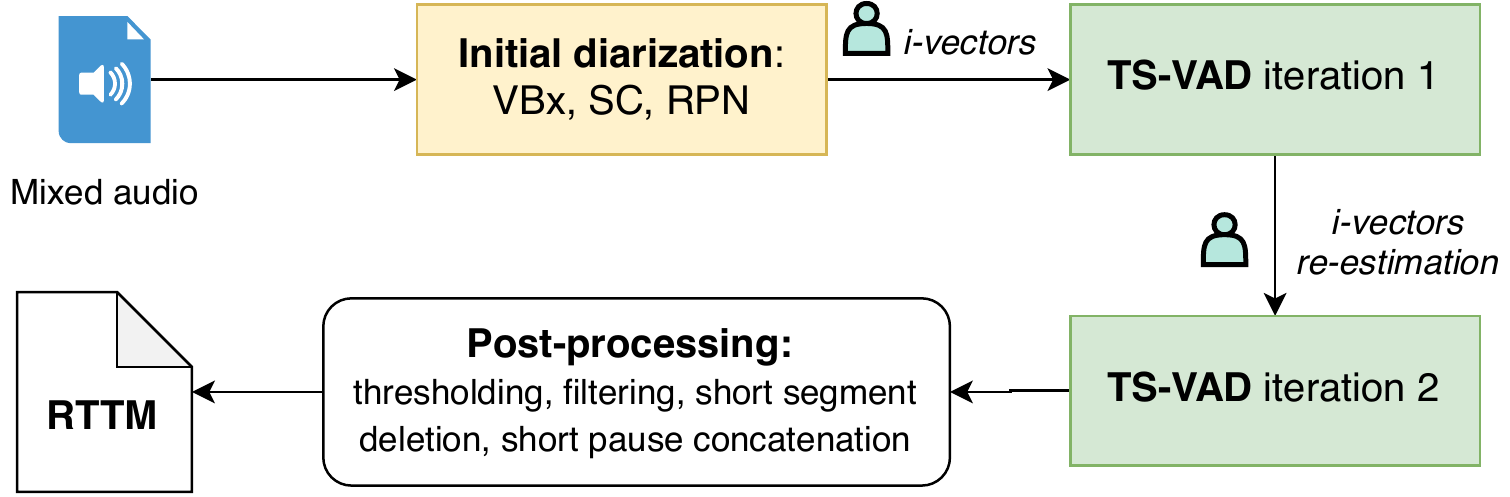}
 \caption{TS-VAD 2-stage decoding pipeline.}
\label{fig:decoding_pipeline}
\end{figure}

As shown in Fig.~\ref{fig:decoding_pipeline}, the general TS-VAD decoding pipeline consists of initialization using an external diarization systems, followed by several iterations of TS-VAD inference, and finally post-processing to convert the network outputs to a sequence of segments (often contained in NIST-style RTTM files). In this paper, we conduct investigations pertaining to several aspects of this decoding pipeline.  

Our key contribution is enabling the TS-VAD model to handle an unknown number of speakers during inference --- in particular, we introduce an inference strategy to deal with a lower number of speakers than was seen during training. 

\subsection{Dataset}
\label{sec:data}

The original TS-VAD model was developed for the CHiME-6 challenge, where each recording consisted of exactly 4 speakers, and using this information was permissible for the participants. Since real-life scenarios may not always adhere to such restrictions, we used meeting-style recordings with varying numbers of speakers for our evaluation. Our evaluation data comprises 3 variants of the LibriCSS dataset~\cite{Chen2020ContinuousSS}, consisting of 2, 5, and 8 speakers, respectively. The dataset contains multi-channel audio recordings of ``simulated conversations,'' generated by mixing test utterances from Librispeech~\cite{Panayotov2015LibrispeechAA}. There are 10 sessions, where each session is approximately one hour long and made up of six 10-minute-long ``mini sessions'' that have different overlap ratios (ranging from 0\% to 40\%). Here, overlap ratio refers to the the fraction of speaking time that contains overlapping speech. The audios were recorded in a regular meeting room using a seven-channel circular microphone array. We selected the first channel of the array for our experiments.

Since LibriCSS does not contain training data, we generated meeting-style audio simulations using training set utterances from Librispeech for training the TS-VAD model. Noise and reverberation were added artificially, and the mixture was created with overlap ratio ranging from 0 to 40\%, similar to the LibriCSS evaluation data. The entire training set comprised approximately 5000 meetings, amounting to 1000 hours of training data.
\subsection{The TS-VAD model}
\label{sec:original}

The TS-VAD model takes conventional speech features (e.g. log Mel filter-banks) as input, along with i-vectors corresponding to the speakers, and predicts per-frame speech activities for all the speakers simultaneously. Formally, given a set of acoustic observations $\mathbf{x} = (\mathbf{x}_1,\ldots,\mathbf{x}_T)$, $\mathbf{x}_t \in \mathbb{R}^D$, and speaker i-vectors $\mathbf{g} = (g_1,\ldots,g_N)$, $g_n \in \mathbb{R}^L$, corresponding to $N$ speakers, the model predicts
\begin{equation}
    \hat{\mathbf{y}} = \text{arg}\max_{\textbf{y}\in [0,1]^{T\times N}} P (\textbf{y}|\mathbf{x},\mathbf{g};\theta),
\end{equation}
where $\mathbf{y}_t^n$ denotes the probability that speaker $n$ is active in frame $t$ of the recording. The probability distribution $P(\textbf{y}|\mathbf{x},\mathbf{g};\theta)$ is modeled using a neural network with parameters $\theta$, and $\theta$ is learned on a training set.

The network architecture consists of 4 convolutional layers which extract acoustic features from raw filter-banks, $\mathbf{x}$. A speaker detection (SD) component comprising 2-layer bidirectional LSTM with projection (BLSTMP) splices these acoustic features along with the i-vectors $\mathbf{g}$ for all the $N$ speakers, and produces $N$ spliced outputs. These are passed to a 1-layer BLSTMP, which finally produces $N$ 2-class outputs corresponding to the speech and silence probabilities for each of the $N$ speakers, namely $\mathbf{y}$.

For training, the number of output nodes $N$ is chosen as the maximum number of speakers in any recording in the training set, which is 8 for our case. The i-vector extractor was trained on Librispeech data with augmented with 3-fold speed perturbation. Since we used simulated training mixtures, we obtained training targets corresponding to the current speaker directly from the forced alignments of the underlying Librispeech utterance.
\begin{figure}[t]
 \centering
 \includegraphics[width=1\linewidth]{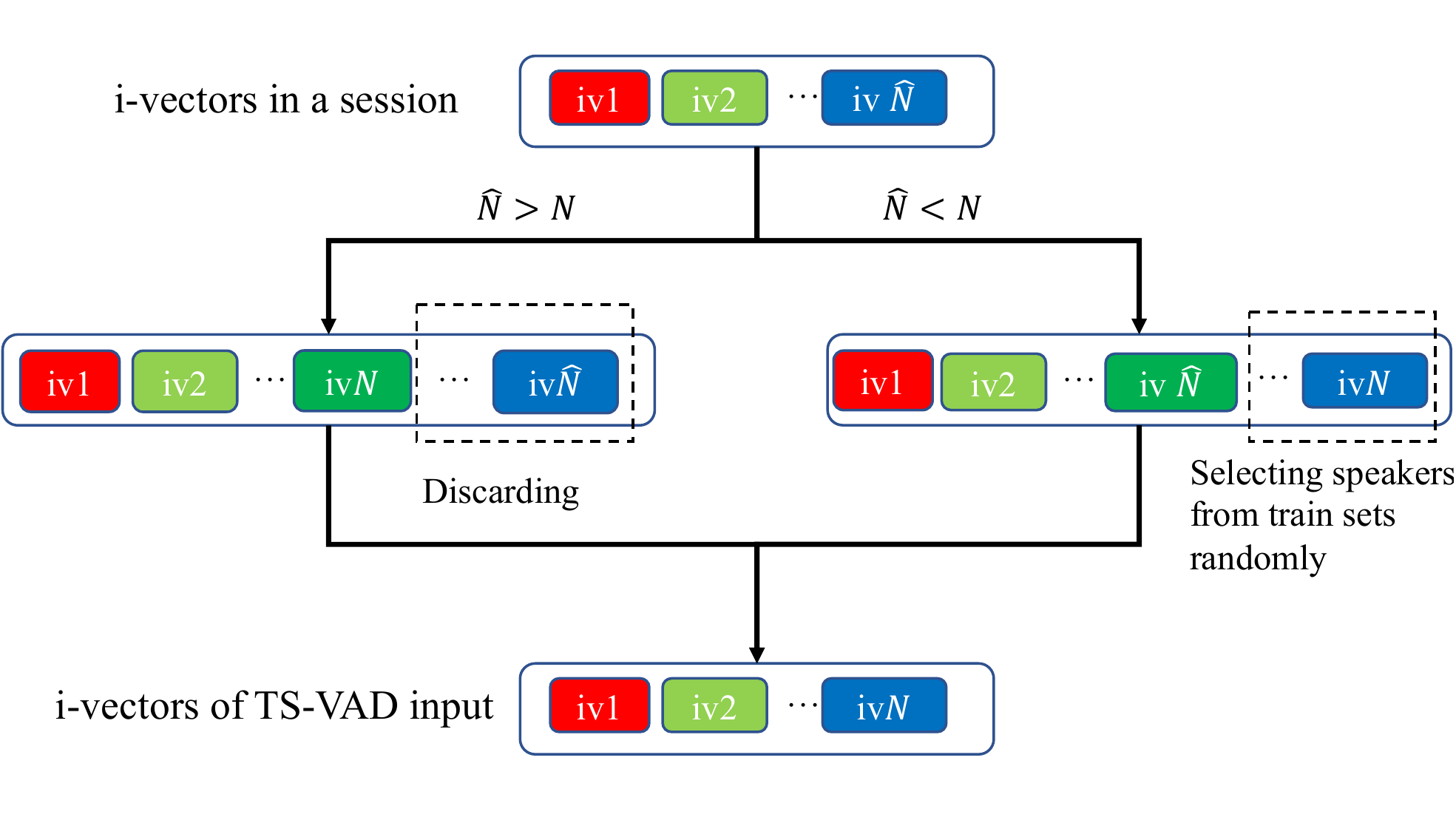}
 \caption{Strategy for handling unknown number of speakers. If the estimated number of speakers $\hat{N}$ exceeds the number of training speakers $N$, we discard the least frequent $\hat{N}-N$ speakers (left branch). If it is less than $N$, we pad the input with $N-\hat{N}$ i-vectors from the training set (right branch).}
\label{fig:ts-vad_input_handling}
\end{figure}

\subsection{Method for handling unknown number of speakers}
\label{sec:proposed}

With the training strategy described above, we obtain a TS-VAD model with a fixed number of output nodes --- which is 8, in our case. At inference time, however, the recording may contain a higher or lower number of speakers. This presents two challenges. First, we need to estimate the number of speakers, say $\hat{N}$, present in the recording. Second, we need to devise a way to use the $N$ output nodes to estimate the frame-level activities of $\hat{N}$ speakers.

Our solution to the first problem is straightforward: we use an existing diarization system to estimate the number of speakers in the recording. Since TS-VAD already requires i-vector estimates to be initialized from another diarization system, this means that this solution does not incur any computational overhead. If this estimate $\hat{N}$ is equal to the number of output nodes $N$, then no further effort is required and the trained model can be directly applied to the recording. Otherwise, there are two possible cases, depending on whether $\hat{N}$ is larger or smaller than $N$. We describe our solutions to both cases below. 

\begin{enumerate}[font={\bfseries},label={Case \arabic*:},wide, labelindent=0pt]

\item If $\hat{N}$ is larger than $N$, we select $N$ out of the estimated speakers who have the longest non-overlapping speaking duration in the initial diarization output, and discard the other speakers. This situation rarely occurs because we chose a larger $N$ for the hole datasets. Even if this happens, the performance loss can be minimized by discarding the short speaker speech.


\item If the estimated number of speakers $\hat{N}$ is smaller than the number of output nodes $N$, we assign $\hat{N}$ of the $N$ output nodes to these ``test'' speakers, and assign the remaining $N - \hat{N}$ nodes to dummy speakers selected from the training set randomly. These dummy training speakers are abandoned at the time of generating the final diarization output.  

The entire procedure is shown in Fig.~\ref{fig:ts-vad_input_handling}.
\end{enumerate}


\section{IMPROVED I-VECTORS ESTIMATION}
\label{sec:i-vectors}

In \cite{Medennikov2020TargetSpeakerVA}, the authors found that the accuracy and iterations required for convergence of the TS-VAD model depended strongly on the diarization system used for initialization. For our second contribution in this paper, we investigated different strategies for initializing the i-vectors used during inference, which we describe in this section.

\subsection{Diarization models for initialization}

We can categorize diarization methods based on whether or not they can assign overlapping speaker segments. Clustering-based diarization \cite{GarciaRomero2017SpeakerDU,Ning2006ASC,Diez2019BayesianHB} is inherently single-speaker, while models such as region proposal networks (RPN)~\cite{Huang2020SpeakerDW} naturally handle overlapping speech. To examine the effect of these different systems for i-vector initialization, we selected the following models.

\begin{enumerate}[wide, labelindent=0pt]

\item \textbf{Spectral clustering (SC)}~\cite{Park2019SpecAugmentAS}: This method consists of a speech activity detection component (SAD) followed by clustering of small subsegment embeddings. We used a similar SAD as that described in \cite{Watanabe2020CHiME6CT}, consisting of a TDNN-Stats based classifier with Viterbi decoding for inference. The speech segments were divided into subsegments with a window size of 1.5s and a stride of 0.75s, and 128-dimensional embeddings were extracted using an x-vector extractor~\cite{Snyder2018XVectorsRD} trained on VoxCeleb data~\cite{Nagrani2017VoxCelebAL}. The subsegment embeddings were scored pair-wise using cosine scoring, and spectral clustering was used to obtain speaker clusters. For estimating the number of speakers, we used an auto-tuning criterion based on $p$-binarization and normalized maximum eigengap~\cite{Park2019SpecAugmentAS}.

\item \textbf{VB-HMM based x-vector clustering (VBx)}~\cite{Diez2019BayesianHB}: Similar to the SC model, we used a TDNN-based SAD followed by subsegment-level embedding extraction using the same x-vector extractor. For clustering, we used VBx, which consists of a Bayesian HMM model. VB inference is used to iteratively refine the soft probabilistic alignment of x-vectors to speakers and re-estimate the speaker specific x-vector distributions. This inference is able to determine the number of speakers in the recording. The PLDA model used for VBx was trained on Librispeech utterances, and the speaker posterior matrix was initialized from the output of an agglomerative hierarchical clustering (AHC) system.

\item \textbf{Region proposal networks (RPN)}~\cite{Huang2020SpeakerDW}: This method combines segmentation and embedding extraction steps into a single neural network, and jointly optimizes them using an objective function that consists of boundary prediction and speaker classification components. The region embeddings are then clustered (using K-means clustering) and a non-maximal suppression is applied. We trained the RPN on simulated meeting-style recordings with partial overlaps generated using utterances from the Librispeech~\cite{Panayotov2015LibrispeechAA} training set.
\end{enumerate}

\cite{2021Integration} gives more details about those diarization methods.

\subsection{Fusion-based initialization strategy}
\label{sec:fusion}

The models mentioned in Section.~\ref{sec:i-vectors} have different (and complementary) strengths. Among clustering-based methods, SC produces an accurate estimate of the number of speakers due to the auto-tuning strategy, while VBx is effective at detecting speaker change with fine granularity. RPN, on the other hand, can detect overlapping speakers. To combine the advantages from these different systems in order to compute more reliable i-vectors, we propose a novel fusion method based on weighted majority voting.

Since diarization outputs may not be in the same label space, we first map them to a common space based on pair-wise overlap duration between speakers, i.e., two speakers from different diarization systems are grouped together if their overlap duration is longer than all other combinations~\footnote{This is similar to the Hungarian method used to compute DER.}. The same metric can be also used for group speakers together from three diarization systems.

\begin{equation}
W^{s}(i)=\sum_{n=1}^{N}{g_n \cdot \text{VAD}_{n}^{s}(i)}, s=1,...,S,
\end{equation}

where $W^s$ is the $s^{th}$ speaker's final weights for i-vector estimation, $N$ is the number of initial diarization systems, and $S$ is the number of speakers. $g_n$ is the weight factor of $n^{th}$ diarization system which satisfies $\sum_{n=1}^{N}{g_n}=1$. In our experiments, we used uniform weights for all systems. $\text{VAD}_s^n$ is an indicator variable which denotes the existence of $s^{th}$ speaker at frame $i$ of the $n^{th}$ diarization system.

\section{EXPERIMENTAL RESULTS}
\label{sec:experiments results}

\subsection{Results for unknown number of speakers}

Table~\ref{tab:2spkand5spk} shows the diarization performance of our TS-VAD model on recordings containing different number of speakers (2, 5, and 8), in terms of diarization error rate (DER) and Jaccard error rate (JER). We trained three different TS-VAD models with different number of output nodes (2, 5, 8) on simulated mixtures containing the respective number of speakers (i.e., the 2spk TS-VAD model was trained on 2-speaker simulated mixtures). We observed a consistent improvement of up to 50\% relative DER with 8spk TS-VAD on all conditions, compared with the SC system that was used for initialization. Furthermore, 8spk TS-VAD achieves even better performance on LibriCSS-2spk and LibriCSS-5spk compared with their custom models. We also used 5spk TS-VAD model to decode LibriCSS-8spk which means losing at least 3 speakers in the final results and the diarization performance drops drastically.This indicates that our heuristic approach for handling unknown number of speakers during inference is effective if we can choose a suitable $N$ for a dataset.

\begin{table}[t]
\centering
\caption{Diarization performance on variants of LibriCSS containing 2, 5, and 8 speakers, in terms of DER and JER. TS-VAD takes i-vectors estimated from the SC output.}
\label{tab:2spkand5spk}
\begin{adjustbox}{max width=\linewidth}
\begin{tabular}{@{}lccccccc@{}}
\toprule
\multicolumn{1}{l}{\multirow{2}{*}{\textbf{Method}}} & \multicolumn{2}{c}{\textbf{LibriCSS-2spk}} & \multicolumn{2}{c}{\textbf{LibriCSS-5spk}} & \multicolumn{2}{c}{\textbf{LibriCSS-8spk}} \\
\cmidrule(r{4pt}){2-3} \cmidrule(r{4pt}){4-5} \cmidrule(l){6-7}
\multicolumn{1}{c}{} & \multicolumn{1}{c}{\textbf{DER}} & \multicolumn{1}{c}{\textbf{JER}} & \multicolumn{1}{c}{\textbf{DER}} & \multicolumn{1}{c}{\textbf{JER}} & \multicolumn{1}{c}{\textbf{DER}} & \multicolumn{1}{c}{\textbf{JER}} \\
\midrule
SC & 21.0 & 20.1 & 19.4 & 20.0 & 14.9 & 20.5 \\
2spk TS-VAD & 12.8 & 12.2 & - & -  & - & - \\
5spk TS-VAD & 12.7 & 12.1 & 11.9 & 12.5 & 42.8 & 45.6 \\
8spk TS-VAD & 12.4 & 11.9 & 11.3 & 12.1 & 7.6 & 11.4 \\
\bottomrule
\end{tabular}
\end{adjustbox}
\end{table}

\subsection{Results for different initializations}

In this section, we present results for our investigation of different initialization strategies for TS-VAD. These experiments were conducted using the 8-speaker LibriCSS dataset, and we report the DERs with a break-down by overlap condition in Table~\ref{tab:mixed_diar}. Note that VBx and SC do not use any prior information about the number of speakers, while RPN uses this oracle information for K-means clustering. We extracted i-vector with non-overlapping speech for overlapping aware initializations like RPN.

\begin{table}[t]
\centering
\caption{Diarization performance on 8-speaker LibriCSS evaluation set, in terms of \% DER. 0S and 0L refer to 0\% overlap with short and long inter-utterance silences, respectively.}
\label{tab:mixed_diar}
\begin{adjustbox}{max width=\linewidth}
\begin{tabular}{@{}llccccccc@{}}
\toprule
\multicolumn{1}{c}{\multirow{2}{*}{\textbf{Method}}} & \multicolumn{1}{c}{\multirow{2}{*}{\textbf{Init.}}} & \multicolumn{6}{c}{\textbf{Overlap ratio in \%}} & \multicolumn{1}{c}{\multirow{2}{*}{\textbf{Avg.}}} \\
\cmidrule(r{4pt}){3-8}
\multicolumn{1}{c}{} & \multicolumn{1}{c}{} & \multicolumn{1}{c}{\textbf{0L}} & \multicolumn{1}{c}{\textbf{0S}} & \multicolumn{1}{c}{\textbf{10}} & \multicolumn{1}{c}{\textbf{20}} & \multicolumn{1}{c}{\textbf{30}} & \multicolumn{1}{c}{\textbf{40}} & \multicolumn{1}{c}{} \\
\midrule
VBx & - & 14.6 & 10.6 & 15.8 & 20.5 & 25.4 & 30.9 & 20.5 \\
SC & - & 11.8 & 9.5 & 12.3 & 15.5 & 18.61 & 18.9 & 14.9 \\
RPN & - & 4.5 & 9.1 & 8.3 & \textbf{6.7} & 11.6 & 14.2 & 9.5 \\
\midrule
TS-VAD & oracle & 2.9 & 4.0 & 5.7 & 5.7 & 8.8 & 7.9 & 6.1 \\
TS-VAD & VBx & 10.3 & 6.8 & 9.3 & 9.0 & 12.6 & 11.6 & 10.0 \\
TS-VAD & SC & 6.0 & \textbf{4.6} & \textbf{6.6} & 7.3 & \textbf{10.3} & \textbf{9.5} & \textbf{7.6} \\
TS-VAD & RPN & \textbf{3.3} & 7.4 & 9.0 & 6.9 &11.7 & 12.3 & 8.9 \\
\bottomrule
\end{tabular}
\end{adjustbox}
\end{table}

We found that TS-VAD improved over all the three initial diarization systems, but this improvement was most prominent for VBx and SC, which cannot handle overlapping segments. The best DERs obtained using SC for initialization are very close to those obtained with i-vectors estimated from oracle segments. Surprisingly, even though the RPN system performed better than VBx and SC, the performance of the TS-VAD model initialized from its output did not achieve the best DERs. 

Next, we evaluated our fusion method described in Section~\ref{sec:fusion}, and the corresponding results are shown in Table~\ref{tab:fusion}. Since TS-VAD initialization with SC provided the best performance for a single system, we retained this system in our fusion, and combined it with the other two initializations, thus providing 3 different fusions. We found that there is small improvement in DER of about 3.9\% relative, compared with the single best TS-VAD system. Besides, by DOVER-Lap \cite{Raj2021DOVERLapAM} of above 3 systems, we got a slight improvement which was better than simply DOVER-Lap \cite{Raj2021DOVERLapAM} of the 3 individual TS-VAD systems from Table~\ref{tab:mixed_diar}.

\begin{table}[t]
\centering
\caption{Diarization performance on LibriCSS evaluation set with different fusion strategies.}
\label{tab:fusion}
\begin{adjustbox}{max width=\linewidth}
\begin{tabular}{@{}lccc@{}}
\toprule
\multicolumn{1}{c}{\textbf{Method}} & \multicolumn{1}{c}{\textbf{Init.}} & \multicolumn{1}{c}{\textbf{DER}} & \multicolumn{1}{c}{\textbf{JER}} \\
\midrule
TS-VAD & SC & 7.6 & 11.4  \\
\midrule
(1) TS-VAD & SC + VBx & 7.5 & 11.1   \\
(2) TS-VAD & SC + RPN  & 7.4 & 11.0 \\
(3) TS-VAD & SC + VBx + RPN  & \textbf{7.3} & \textbf{11.0} \\
\midrule
DOVER-Lap of TS-VAD Init. with (VBx, SC, RPN) & - & 7.3 & 11.2 \\
DOVER-Lap of (1,2,3) & - & 7.1 & 10.8 \\
\bottomrule
\end{tabular}
\end{adjustbox}
\end{table}

\subsection{Impact on ASR performance}

We also evaluated the impact of the TS-VAD diarization system on downstream ASR performance. For this, we built a hybrid HMM-DNN system following the Kaldi~\cite{Povey2011TheKS} Librispeech recipe. In Table~\ref{tab:mixed_asr}, we present the ASR performance in terms of concatenated minimum-permutation word error rates (cpWERs)~\cite{Watanabe2020CHiME6CT}. The cpWER is computed by concatenating all the utterances of a speaker in the reference and hypothesis, scoring all speaker pairs, and then finding the speaker permutation that minimizes the total WER. We compared the cpWERs obtained using 3 different segmentation methods: oracle, SC, and 8-spk TS-VAD, where the 8-spk TS-VAD model was initialized using the SC output. We found that the going from SC to TS-VAD resulted in a significant cpWER improvement from 32.2\% to 25.8\% for LibriCSS-8spk set. This was very close to the cpWER obtained using oracle segments, which was 23.1\%. Similar improvements were observed using the 8spk TS-VAD on both LibriCSS-2spk and LibriCSS-5spk compared with the SC system. However, the cpWER on such conditions was still high, indicating that better ASR models, or an external speech separation module may be required to satisfactorily transcribe such recordings.

\begin{table}[t]
\centering
\caption{ASR performance on mixed LibriCSS evaluation set with different diarization segments, in terms of \% WER.}
\label{tab:mixed_asr}
\begin{adjustbox}{max width=\linewidth}
\begin{tabular}{cccc}
\toprule
\textbf{Segments} & \textbf{LibriCSS-2spk} & \textbf{LibriCSS-5spk} & \textbf{LibriCSS-8spk} \\
\midrule
Oracle & 26.3 & 25.7 & 23.1 \\
SC & 35.7 & 34.5 & 32.2 \\
8spk TS-VAD & 29.4 & 27.9 & 25.8 \\
\bottomrule
\end{tabular}
\end{adjustbox}
\end{table}

\section{Conclusion}

We adapted TS-VAD to the diarization of multi-speaker conversations, which provided state-of-the-art results in meeting scenarios where the number of speakers is unknown. Through experiments on LibriCSS variants comprising 2, 5, and 8 speakers, we showed the efficacy of our proposed solution. Notably, an 8-spk TS-VAD model outperformed customized models built for smaller number of speakers. We also proposed a simple strategy to estimate the input i-vectors for TS-VAD using multi-initial diarization results, which gave us further improvements. Our investigations on downstream ASR performance suggested that while we can get close to oracle segmentation performance using TS-VAD, a separation module may indeed be necessary, since overlapping speech is a bottleneck for the ASR system. In future work, we will investigate diarization based separation methods for ASR to avoid this issue.

 \section{Acknowledgments}
The work reported here was started at JSALT 2020, with support from Microsoft, Amazon, and Google. We thank Tianyan Zhou, Xiaofei Wang, and Zhong Meng for their contributions for collecting LibriCSS 2spk and 5spk data. This work was supported by the Strategic Priority Research Program of Chinese Academy of Sciences under Grant No. XDC08050200

\bibliographystyle{IEEEtran}

\bibliography{refs}

\begin{thebibliography}{10}
\providecommand{\url}[1]{#1}
\csname url@samestyle\endcsname
\providecommand{\newblock}{\relax}
\providecommand{\bibinfo}[2]{#2}
\providecommand{\BIBentrySTDinterwordspacing}{\spaceskip=0pt\relax}
\providecommand{\BIBentryALTinterwordstretchfactor}{4}
\providecommand{\BIBentryALTinterwordspacing}{\spaceskip=\fontdimen2\font plus
\BIBentryALTinterwordstretchfactor\fontdimen3\font minus
  \fontdimen4\font\relax}
\providecommand{\BIBforeignlanguage}[2]{{%
\expandafter\ifx\csname l@#1\endcsname\relax
\typeout{** WARNING: IEEEtran.bst: No hyphenation pattern has been}%
\typeout{** loaded for the language `#1'. Using the pattern for}%
\typeout{** the default language instead.}%
\else
\language=\csname l@#1\endcsname
\fi
#2}}
\providecommand{\BIBdecl}{\relax}
\BIBdecl

\bibitem{Mir2012SpeakerDA}
X.~A. Mir{\'o}, S.~Bozonnet, N.~W.~D. Evans, C.~Fredouille, G.~Friedland, and
  O.~Vinyals, ``Speaker diarization: A review of recent research,'' \emph{IEEE
  Transactions on Audio, Speech, and Language Processing}, vol.~20, pp.
  356--370, 2012.

\bibitem{Tranter2006AnOO}
S.~Tranter and D.~A. Reynolds, ``An overview of automatic speaker diarization
  systems,'' \emph{IEEE Transactions on Audio, Speech, and Language
  Processing}, vol.~14, pp. 1557--1565, 2006.

\bibitem{GarciaRomero2017SpeakerDU}
D.~Garcia-Romero, D.~Snyder, G.~Sell, D.~Povey, and A.~McCree, ``Speaker
  diarization using deep neural network embeddings,'' \emph{2017 IEEE
  International Conference on Acoustics, Speech and Signal Processing
  (ICASSP)}, pp. 4930--4934, 2017.

\bibitem{Sun2018SpeakerDW}
L.~Sun, J.~Du, C.~Jiang, X.~Zhang, S.~He, B.~Yin, and C.-H. Lee, ``Speaker
  diarization with enhancing speech for the first dihard challenge,'' in
  \emph{INTERSPEECH}, 2018.

\bibitem{Dehak2011FrontEndFA}
N.~Dehak, P.~Kenny, R.~Dehak, P.~Dumouchel, and P.~Ouellet, ``Front-end factor
  analysis for speaker verification,'' \emph{IEEE Transactions on Audio,
  Speech, and Language Processing}, vol.~19, pp. 788--798, 2011.

\bibitem{Variani2014DeepNN}
E.~Variani, X.~Lei, E.~McDermott, I.~Lopez-Moreno, and J.~Gonzalez-Dominguez,
  ``Deep neural networks for small footprint text-dependent speaker
  verification,'' \emph{ICASSP}, pp. 4052--4056, 2014.

\bibitem{Snyder2018XVectorsRD}
D.~Snyder, D.~Garcia-Romero, G.~Sell, D.~Povey, and S.~Khudanpur, ``X-vectors:
  Robust {DNN} embeddings for speaker recognition,'' \emph{2018 IEEE
  International Conference on Acoustics, Speech and Signal Processing
  (ICASSP)}, pp. 5329--5333, 2018.

\bibitem{Bozonnet2010TheLR}
S.~Bozonnet, N.~Evans, and C.~Fredouille, ``The lia-eurecom rt'09 speaker
  diarization system: Enhancements in speaker modelling and cluster
  purification,'' \emph{2010 IEEE International Conference on Acoustics, Speech
  and Signal Processing}, pp. 4958--4961, 2010.

\bibitem{Sell2015DiarizationRI}
G.~Sell and D.~Garcia-Romero, ``Diarization resegmentation in the factor
  analysis subspace,'' \emph{2015 IEEE International Conference on Acoustics,
  Speech and Signal Processing (ICASSP)}, pp. 4794--4798, 2015.

\bibitem{boakye2008overlapped}
K.~Boakye, B.~Trueba-Hornero, O.~Vinyals, and G.~Friedland, ``Overlapped speech
  detection for improved speaker diarization in multiparty meetings,'' in
  \emph{2008 IEEE International Conference on Acoustics, Speech and Signal
  Processing}.\hskip 1em plus 0.5em minus 0.4em\relax IEEE, 2008, pp.
  4353--4356.

\bibitem{Huijbregts2009SpeechOD}
M.~Huijbregts, D.~A. van Leeuwen, and F.~de~Jong, ``Speech overlap detection in
  a two-pass speaker diarization system,'' in \emph{INTERSPEECH}, 2009.

\bibitem{Hagerer2017EnhancingLR}
G.~Hagerer, V.~Pandit, F.~Eyben, and B.~W. Schuller, ``Enhancing {LSTM}
  {RNN}-based speech overlap detection by artificially mixed data,'' in
  \emph{Semantic Audio}, 2017.

\bibitem{Kunesov2019DetectionOO}
M.~Kunesov{\'a}, M.~Hr{\'u}z, Z.~Zaj{\'i}c, and V.~Radov{\'a}, ``Detection of
  overlapping speech for the purposes of speaker diarization,'' in
  \emph{SPECOM}, 2019.

\bibitem{Bullock2019OverlapawareDR}
L.~Bullock, H.~Bredin, and L.~P. Garc{\'i}a-Perera, ``Overlap-aware
  diarization: resegmentation using neural end-to-end overlapped speech
  detection,'' \emph{ArXiv}, vol. abs/1910.11646, 2019.

\bibitem{Fujita2020EndtoEndND}
Y.~Fujita, S.~Watanabe, S.~Horiguchi, Y.~Xue, and K.~Nagamatsu, ``End-to-end
  neural diarization: Reformulating speaker diarization as simple multi-label
  classification,'' \emph{ArXiv}, vol. abs/2003.02966, 2020.

\bibitem{Huang2020SpeakerDW}
Z.~Huang, S.~Watanabe, Y.~Fujita, P.~Garc{\'i}a, Y.~Shao, D.~Povey, and
  S.~Khudanpur, ``Speaker diarization with region proposal network,''
  \emph{ICASSP 2020 - 2020 IEEE International Conference on Acoustics, Speech
  and Signal Processing (ICASSP)}, pp. 6514--6518, 2020.

\bibitem{Medennikov2020TargetSpeakerVA}
I.~Medennikov, M.~Korenevsky, T.~Prisyach, Y.~Y. Khokhlov, M.~Korenevskaya,
  I.~Sorokin, T.~V. Timofeeva, A.~Mitrofanov, A.~Andrusenko, I.~Podluzhny,
  A.~Laptev, and A.~Romanenko, ``Target-speaker voice activity detection: a
  novel approach for multi-speaker diarization in a dinner party scenario,''
  \emph{ArXiv}, vol. abs/2005.07272, 2020.

\bibitem{Watanabe2020CHiME6CT}
S.~Watanabe, M.~Mandel, J.~Barker, and E.~Vincent, ``{CHiME-6} challenge:
  Tackling multispeaker speech recognition for unsegmented recordings,''
  \emph{ArXiv}, vol. abs/2004.09249, 2020.

\bibitem{molkov2017SpeakerAwareNN}
K.~Žmol{\'i}kov{\'a}, M.~Delcroix, K.~Kinoshita, T.~Higuchi, A.~Ogawa, and
  T.~Nakatani, ``Speaker-aware neural network based beamformer for speaker
  extraction in speech mixtures,'' in \emph{INTERSPEECH}, 2017.

\bibitem{Wang2019VoiceFilterTV}
Q.~Wang, H.~Muckenhirn, K.~Wilson, P.~Sridhar, Z.~Wu, J.~Hershey, R.~A.
  Saurous, R.~J. Weiss, Y.~Jia, and I.~Lopez-Moreno, ``Voicefilter: Targeted
  voice separation by speaker-conditioned spectrogram masking,'' in
  \emph{INTERSPEECH}, 2019.

\bibitem{Ding2019PersonalVS}
S.~Ding, Q.~Wang, S.~yiin Chang, L.~Wan, and I.~Lopez-Moreno, ``Personal vad:
  Speaker-conditioned voice activity detection,'' \emph{ArXiv}, vol.
  abs/1908.04284, 2019.

\bibitem{Chen2020ContinuousSS}
Z.~Chen, T.~Yoshioka, L.~Lu, T.~Zhou, Z.~Meng, Y.~Luo, J.~Wu, and J.~Li,
  ``Continuous speech separation: Dataset and analysis,'' \emph{ICASSP 2020 -
  2020 IEEE International Conference on Acoustics, Speech and Signal Processing
  (ICASSP)}, pp. 7284--7288, 2020.

\bibitem{Raj2021MultiSC}
D.~Raj, Z.~Huang, and S.~Khudanpur, ``Multi-class spectral clustering with
  overlaps for speaker diarization,'' In submission.

\bibitem{Panayotov2015LibrispeechAA}
V.~Panayotov, G.~Chen, D.~Povey, and S.~Khudanpur, ``Librispeech: An {ASR}
  corpus based on public domain audio books,'' \emph{2015 IEEE International
  Conference on Acoustics, Speech and Signal Processing (ICASSP)}, pp.
  5206--5210, 2015.

\bibitem{Ning2006ASC}
H.~Ning, M.~Liu, H.~Tang, and T.~S. Huang, ``A spectral clustering approach to
  speaker diarization,'' in \emph{INTERSPEECH}, 2006.

\bibitem{Diez2019BayesianHB}
M.~Diez, L.~Burget, S.~Wang, J.~Rohdin, and J.~Cernock{\'y}, ``Bayesian {HMM}
  based x-vector clustering for speaker diarization,'' in \emph{INTERSPEECH},
  2019.

\bibitem{Park2019SpecAugmentAS}
D.~S. Park, W.~Chan, Y.~Zhang, C.-C. Chiu, B.~Zoph, E.~D. Cubuk, and Q.~V. Le,
  ``Specaugment: A simple data augmentation method for automatic speech
  recognition,'' in \emph{INTERSPEECH}, 2019.

\bibitem{Nagrani2017VoxCelebAL}
A.~Nagrani, J.~S. Chung, and A.~Zisserman, ``Voxceleb: A large-scale speaker
  identification dataset,'' \emph{ArXiv}, vol. abs/1706.08612, 2017.

\bibitem{2021Integration}
D.~Raj, P.~Denisov, Z.~Chen, H.~Erdogan, and J.~R. Hershey, ``Integration of
  speech separation, diarization, and recognition for multi-speaker meetings:
  System description, comparison, and analysis,'' in \emph{2021 IEEE Spoken
  Language Technology Workshop (SLT)}, 2021.

\bibitem{Raj2021DOVERLapAM}
D.~Raj, L.~P. Garc{\'i}a-Perera, Z.~Huang, S.~Watanabe, D.~Povey, A.~Stolcke,
  and S.~Khudanpur, ``{DOVER-Lap}: A method for combining overlap-aware
  diarization outputs,'' \emph{2021 IEEE Spoken Language Technology Workshop
  (SLT)}, pp. 881--888, 2021.

\bibitem{Povey2011TheKS}
D.~Povey, A.~Ghoshal, G.~Boulianne, L.~Burget, O.~Glembek, N.~K. Goel,
  M.~Hannemann, P.~Motl{\'i}cek, Y.~Qian, P.~Schwarz, J.~Silovsk{\'y},
  G.~Stemmer, and K.~Vesel{\'y}, ``The kaldi speech recognition toolkit,''
  2011.

\end{thebibliography}

\end{document}